\documentclass[aps,pra,epsfigure,twocolumn,showpacs]{revtex4}

\usepackage{epsfig}

\usepackage{dcolumn}

\usepackage{bm}

\usepackage{graphicx}

\usepackage{amsmath}

\usepackage{latexsym}

\usepackage{amsfonts}

\usepackage{amssymb}

\usepackage{array}

\newcommand{\hf}{\frac{1}{2}}

\newcommand{\numop}{\hat{a}^{\dagger}\hat{a}}

\newcommand{\exg}{\left|e\right>\left<g\right|}

\newcommand{\gxe}{\left|g\right>\left<e\right|}

\newcommand{\ea}{\left|e\right>}

\newcommand{\ga}{\left|g\right>}

\newcommand{\ket}[1]{|#1\rangle}

\newcommand{\bra}[1]{\langle#1|}

\newcommand{\braket}[2]{\langle#1|#2\rangle}

\begin{document}


\title{Concentration and purification of entanglement for qubit systems with ancillary cavity fields}

\author{C. D. Ogden,  M. Paternostro, and M. S. Kim}

\affiliation{School of Mathematics and Physics, Queen's University, Belfast BT7 1NN, United Kingdom}

\begin{abstract}

We propose schemes for entanglement concentration and purification for qubit systems encoded in flying atomic pairs. We use a cavity-quantum electrodynamics setting as the paradigmatic scenario within which our proposals can be implemented. Maximally entangled pure states of qubits can be produced as a result of our protocols. In particular, the concentration protocol yields Bell states with the largest achievable theoretical probability while the purification scheme produces arbitrarily pure Bell states. The requirements for the implementation of these protocols are modest, within the state of the art, and we address all necessary steps in two specific set-ups based on experimentally mature microwave technology.



\end{abstract}

\maketitle

\section{Introduction}

Besides the attention that modern physics pays to entanglement in virtue of the role it plays in testing fundamental properties of quantum systems, there is an enormous interest in developing strategies and protocols allowing for the manipulation of these intimately quantum mechanical correlations between discrete as well as continuous variable systems.  Even though its role in computational processes is not yet fully understood and is the subject of intriguing investigations, it is nowadays recognized that the ability of manipulating entanglement is crucial for many tasks in quantum communication and, in general, quantum information processing (QIP).

Over the past two decades, we have witnessed many proposals for the generation of entangled pairs of quantum systems, in various settings~\cite{nielsenchuang}. For instance, it has become a routine procedure in linear optics QIP with discrete variables, in which parametric down conversion is used~\cite{zeilinger} and in its continuous variable version, where two-mode squeezed states are effortlessly produced by non-degenerate optical parametric amplifiers. In different contexts, entanglement of pairs of atoms~\cite{hagley}, trapped ions~\cite{ionbyte}, superconducting devices~\cite{nakamura} and atom-photon systems~\cite{duanmonroe} has already been demonstrated.

Regardless of the efficiency with which it can be generated, entanglement is undoubtedly a fragile resource, susceptible to the detrimental effects of communication channels and environmental interactions that result in a degradation of the desired ``quality" of the quantum resource. This is particularly true in a scenario of distributed QIP, where the exchange of high-quality entangled states (which should be as pure as possible and carrying a full ebit of quantum correlations) is necessary in order to realize perfectly efficient computational protocols. This has lead to the development of strategies to increase the entanglement and purity of mixed, non-maximally entangled states \cite{bennett1,bennett2}. So far, experimental realizations of entanglement purification schemes have been limited to linear optics~\cite{J-W. Pan  Nature 423 417 (2003), Imoto Nature 421 343 (2003), Kwiat Nature 409 1014 (2001)}
and trapped-ions~\cite{Reichle Nature 443 838 (2006)}. Though lacking such a demonstration, microwave cavity-quantum electrodynamics (cQED) holds great promise as a medium for QIP. Strong coupling between atoms and field modes allows for the creation~\cite{haroche} and distribution~\cite{noi} of entanglement, and the performance of quantum logic operations~\cite{M. s. Zubairy et al pra 68 033820 (2003)}. The new high finesse cavities~\cite{kuhr} demonstrated by Kuhr {\em et al.} improve the feasibility a system of distributed QIP employing atoms as flying qubits, and cavity field modes as relatively long-lived static qubits.  This system would clearly be enhanced by methods for improving degraded entanglement.  
 
Suggestions have been made for cQED purification schemes that recover entanglement only from decoherence due to cavity decay ~\cite{J. L. Romero et al PRA 65 052319 (2002)}, or that rely on repeated interactions of the cavity modes with a single flying atom~\cite{compagno pra 2004}.  Here, we propose schemes for the concentration of pure-state entanglement, and the distillation of pure-state entanglement from Werner states (assuming a sufficiently large fraction of any Bell state) and maximally entangled mixed states (MEMS)~\cite{munro}.  Our protocols rely only on single-atom operations and measurements, and could be performed using current technology.





The paper is organized as follows. In Sec.~\ref{concentration} we introduce the interaction model used throughout this investigation and describe the protocol for the concentration of entanglement that represents the first part of our study. This probabilistic entanglement concentration is accompanied by a simultaneous interconversion from flying-qubit to stationary-qubit entanglement. We prove that a sufficiently long interaction time allows for the concentration of entanglement to the maximal value of a full ebit with the highest probability being theoretically allowed. The scheme requires resources that can be prepared off-line and is extremely close to ideal efficiency for values of the experimentally relevant parameters attainable with present-day technology. In Sec.~\ref{purification} we analyze an entanglement purification protocol that, when operating on mixed Bell-diagonal states, extracts a progressively more entangled and pure two-qubit state. The scheme is probabilistic and requires information gained on the population of the cavity fields. We demonstrate the wide applicability of our scheme by addressing the purification of two interesting classes of mixed states, namely Werner states and maximally entangled mixed states of two qubits. Sec.~\ref{practical} is devoted to the analysis of the feasibility of the schemes we propose in microwave cavity- and circuit-QED set-ups. We highlight how the recent achievements in this field allow for a foreseeable implementation of our proposals. Moreover, we discuss the possibilities offered by superconducting devices integrated in planar resonators as alternative scenarios in which our protocols can be realized. Finally, Sec.~\ref{conclusions} summarizes our results.

\section{Concentration Scheme}

\label{concentration}

Many problems involving the interaction of spin-like two-level systems with single-mode bosonic systems can be modelled with an effective dipole-coupling formalism in which a spin operator (generally proportional to the $\hat{\sigma}_x$ Pauli spin operator) is coupled to the electric (or the magnetic) part of an electromagnetic field.  This is, evidently, the case for neutral atoms or quantum dots coupled to optical or microwave fields~\cite{hagley,rauschenbeutel,varie}. On the other hand, this description holds also for a system consisting of a Cooper-pair box (in a superconducting-quantum-interference device (SQUID) configuration and in the charge regime~\cite{schon}) integrated into a planar stripline resonator~\cite{schoelkopf}, a setting we generally refer to as {\it circuit-QED}. At the charge degeneracy point, an effective dipole moment operator for the SQUID can be written, whose amplitude is proportional to the excess charge in the SQUID island~\cite{schoelkopf}. In what follows, in order to fix the ideas and introduce the general formalism employed throughout our study, we shall use language and terminology that are typical of cavity-QED, and we refer explicitly to a scheme of Rydberg atoms interacting with microwave cavities. In Sec.~\ref{practical} we extensively assess the details of oimplementation in both cavity- and circuit-QED.

Let us consider a two-level atom with ground and excited states $\ket{g}$ and $\ket{e}$, respectively. The corresponding transition frequency is labelled $\omega_o$. The atom interacts with the field mode of a cavity of frequency $\omega_f$, described by the bosonic annihilation (creation) operator $\hat{a}$ ($\hat{a}^{\dag}$). Within the dipole-coupling interaction assumed here, the total Hamiltonian of this atom-field system is
\begin{equation}
\label{full JC ham}
\hat{H}= \hbar\omega_o\hat{\sigma}_z+ \hbar\omega_f\numop+ \hbar\lambda\left( \hat{\sigma}_+ + \hat{\sigma}_- \right) \left ( \hat{a} + \hat{a}^{\dagger} \right),
\end{equation}
where $\hat{\sigma}_+=\exg$ and $\hat{\sigma}_- =\gxe$ are the atomic ladder operators and $\lambda$ is the coupling strength of the interaction.  Employing the rotating wave approximation~\cite{gerryknight}, assuming atom-field resonance ({\it i.e.} $\omega_0\simeq{\omega_f}$) and entering the interaction picture with respect to the free energy $\hat{H}_0=\hbar\omega_o\hat{\sigma}_z+ \hbar\omega_o\numop$, the coupling reduces to the standard Jaynes-Cummings model~\cite{jc} 

\begin{equation}
\label{int JC ham}
\hat{H}_{int} =\hbar\lambda\left( \hat{\sigma}_+ \hat{a} + \hat{\sigma}_-\hat{a}^{\dagger}\right).
\end{equation}

The dynamics arising from this Hamiltonian~(\ref{int JC ham}) are described by the following time-evolution operator, written in the atomic basis $\{\ket{e},\ket{g}\}$~\cite{phoenix}
\begin{equation}
\label{timeevolve}
\hat U(t) =
\left[
\begin{matrix} 
      \cos{ 
      (
         \lambda t\sqrt{\hat n + 1}
      )}
   &
      -i\hat{a}
      \frac{
         \sin{
    (
       \lambda t\sqrt{\hat n}
    )}
      }{
    {\sqrt{\hat n}}
      }
   \\
      -i\hat{a}^\dag
      \frac{
         \sin{
    (
       \lambda t\sqrt{\hat n + 1}
    )}
      }{
         {\sqrt{\hat n + 1}}
      }
   &
      \cos{
      (
         \lambda t\sqrt{\hat n}
      )}
   \end{matrix}
\right],
\end{equation}
where $\hat n=\hat{a}^{\dag}\hat{a}$ is the photon-number operator of the cavity field. 

Even though this simple interaction model has been extensively studied in recent years, the large variety of effects for which it is responsible and its vast applicability to problems of QIP make it worth our consideration. Here, we give a simple, yet interesting example of its applicability to the problem of entanglement concentration. 

Consider a non-maximally entangled pure state of two atoms, labelled $1$ and $2$:

\begin{equation}
\label{non max phiplus}
\ket{\Psi_o}_{12} = 
\alpha\ket{ee}_{12} + \beta\ket{gg}_{12}.
\end{equation}

where $\alpha$ is real and $\beta = \sqrt{1-\alpha^2}$. The assumption of this form of the input state does not affect the generality of our approach. (Other input states may be used if converted into the form~(\ref{non max phiplus}) , which can be achieved by using Ramsey zones to perform single-qubit rotations.)  This state can be created using a variant of the scheme suggested in Refs.~\cite{accumulation}, which is able to establish an arbitrary amount of entanglement in the pure state of two {\it preparatory} cavities $c1$ and $c2$, {\it i.e.} a state of the form $\cos(\pi\tau)\ket{01}_{c1,c2}+\sin(\pi\tau)\ket{10}_{c1,c2}$. Atoms 1 and 2 are prepared in their ground states, and each is passed through one of the two cavities.  By arranging a quarter-Rabi cycle between atom $j$ and the field of cavity $cj$ ($j=1,2$), and using a single Ramsey zone~\cite{haroche} to perform a $\sigma_z$ rotation of angle $\pi$ on one of the atoms, we will transfer the entanglement from the cavities to the atoms.  Alternatively, one can use a single cavity sequentially crossed by atoms $1$ and $2$ as in Ref.~\cite{rauschenbeutel}. Of course, the preparation of the input state can be considered as an off-line step that should not be accounted among the requirements of the scheme we address.

We present two versions of our concentration protocol; a symmetric scheme, in which $\alpha$ can take any value in the range $[-1,1]$, and an asymmetric scheme in which we require $\alpha > \beta$.

\subsection{Symmetric Concentration Scheme} 

\begin{figure}[t]
\centerline{\includegraphics[width=0.4\textwidth]{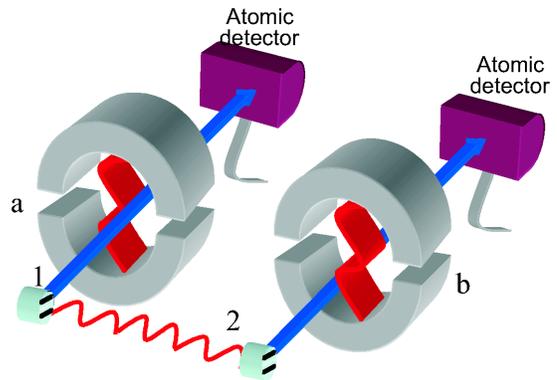}}
\caption{Entanglement concentration scheme. Two cavities act as stationary qubits interacting with flying atomic pairs previously prepared in a non-maximally entangled state. Local cavity-atom interactions followed by atomic-state detection probabilisticly transfer one ebit of entanglement to the cavity field modes. For properly chosen interaction times, the probability of obtaining an ebit is arbitrarily close to the optimal theoretical value.}
\label{fig:fig0}
\end{figure}

We allow qubits $1$ and $2$ to interact with the field of cavities $a$ and $b$ respectively, which are spatially separated with each prepared in a single-photon state. A sketch of the described setup is presented in Fig.~\ref{fig:fig0}. The initial state of the overall system is, therefore, $\ket{\Psi_o}_{12}\otimes\ket{11}_{ab}$ which evolves according to the global time-propagator $\hat{U}_{1a}(t_1)\otimes\hat{U}_{2b}(t_2)$, where $\hat{U}_{j\mu}(t_j)$ is given by the matrix expression Eq.~(\ref{timeevolve}) with the replacement $\hat{a},\hat{a}^\dag\rightarrow\hat{a}_j,\hat{a}^{\dag}_j$ ($j=a,b$) and atomic label $\mu=1,2$.

For the sake of simplicity, we set the interaction time $t_{1}=t_2=t$, which results in the evolved state

\begin{equation}
\label{evoluto}
\begin{aligned}
\ket{\Psi_{af}}_{12ab}&=\frac{1}{2}[
\ket{ee,\psi_{ee}}_{12ab} - \ket{gg,\psi_{gg}}_{12ab}\\
&-i\ket{eg,\psi_{eg}}_{12ab} -i \ket{ge,\psi_{ge}}_{12ab}].
\end{aligned}
\end{equation}
In this equation, we have introduced the normalized two-mode states
\begin{equation}
\label{ensemble}
\begin{aligned}
\ket{\psi_{ee}}_{ab} &={\cal N}_{ee}[
   \alpha \cos^2{(\sqrt{2}\lambda t)}\ket{11}_{ab}
   -
   \beta \sin^2{( \lambda t )}\ket{00}_{ab}],\\
\ket{\psi_{eg}}_{ab} &={\cal N}_{eg}[
   {\alpha}{} \cos{(2\sqrt{2}\lambda t)} 
      \ket{12}_{ab}
   +
   {\beta}{} \sin{\left(2\lambda t \right)}\ket{01}_{ab}],\\
\ket{\psi_{ge}}_{ab} & ={\cal N}_{ge}[
   {\alpha}{} \cos{(2\sqrt{2}\lambda t)} 
    \ket{21}_{ab}
   +
   {\beta}{}\sin{(2\lambda t)}\ket{10}_{ab}],\\
\ket{\psi_{gg}}_{ab} & ={\cal N}_{gg}[
   \alpha \sin^2{(\sqrt{2}\lambda t)}\ket{22}_{ab}
   -
   \beta \cos^2{( \lambda t)}\ket{11}_{ab}]
\end{aligned}
\end{equation}
For the sake of mathematical convenience, we shall consider the case of $0<\alpha<\beta$.  These equations show that the information initially contained in the input two-atom state has been transferred to cavity fields in the subspace $\{\ket{0}_j,\ket{1}_j,\ket{2}_j\}$ of the relevant infinite-dimensional Hilbert space. To complete the entanglement concentration process, we measure the state of each atom in the $\hat{\sigma}_z$-eigenbasis $\{\ea,\ga\}$. After postselecting the events corresponding to both atoms being found in $\ket{e}$, this operation results in the projection of the cavity fields onto $\ket{\psi_{ee}}_{ab}$
with probability
\begin{equation}
\label{prob}
P( \ket{\psi_{ee}}) = 
\alpha^2[\cos{(\sqrt{2}\lambda t )}]^4
+
\beta^2 [\sin{( \lambda t)}]^4.
\end{equation}
If the interaction time was selected such that 
\begin{equation}
\label{condn for bell state}
\alpha[\cos(\sqrt{2}\lambda t )]^2 = \beta[\sin( \lambda t)]^2
\end{equation}
we produce a two-mode state with a full ebit of entanglement.  Therefore, we have probabilistically concentrated the initial flying-qubit entanglement by transferring it to two stationary qubits, here embodied by the cavity field modes. This full ebit can, with unit probability, be transferred to two flying atoms by arranging local interactions with the cavity fields for a time such that the transformations $\ket{g,0}_{atom,cavity}\rightarrow\ket{g,0}_{atom,cavity},\,\ket{g,1}_{atom,cavity}\rightarrow\ket{e,0}_{atom,cavity}$ are realized. 


Let us now consider whether or not this concentration scheme is ideal.
%
%
To this end, we use the tool provided by the {\em entanglement of single pair purification} (ESPP), $E_S$~\cite{s. bose et al pra 60 194 (1999)}. This quantity, which is shown to be a measure of entanglement in Ref.~\cite{s. bose et al pra 60 194 (1999)}, is defined as the maximum probability with which a non-maximally entangled qubit pair can be converted to a full-ebit state using local operations and classical communication (LOCC). It is equal to $2|c_s|^2$, where $c_s$ is the smallest Schmidt coefficient in the state under investigation~\cite{s. bose et al pra 60 194 (1999)}. The input state~(\ref{non max phiplus}) has $E_S = 2\alpha^2$. The optimality of the concentration protocol is evaluated as follows: As ESPP is a good measure of entanglement, it cannot increase, on average, under application of LOCC~\cite{vlatko}. Therefore, the protocol for entanglement concentration is said to be optimal if it conserves ESPP.

As before the measurement of the atomic pair we have an {\it a priori} ensemble of states given by Eqs.~(\ref{ensemble}), each obtained with a precise probability, the object of our interest will be the average entanglement of the ensemble. After measuring the state of the atomic pair, the cavity modes have the average ESPP

\begin{equation}
\begin{aligned}
\left< E_S \right>_f
&= 
p\left( \ket{\psi_{ee}}\right) 
E_S\left( \ket{\psi_{ee}}\right)
+
p\left( \ket{\psi_{eg}}\right) 
E_S\left( \ket{\psi_{eg}}\right)\\
&+
p\left( \ket{\psi_{ge}}\right) 
E_S\left( \ket{\psi_{ge}}\right)
+
p\left( \ket{\psi_{gg}}\right) 
E_S\left( \ket{\psi_{gg}}\right),
\end{aligned}
\end{equation}

where $p\left( \ket{\psi_{ij}}\right)$ is the probability of projecting the field modes onto the state $\ket{\psi_{ij}}$ ($i,j=e,g$), which has ESPP $E_S(\ket{\psi_{ij}})$. 

\begin{figure}[t]
\centerline{\bf{(a)}\hskip3cm\bf{(b)}}
\centerline{\includegraphics[width=0.18\textwidth]{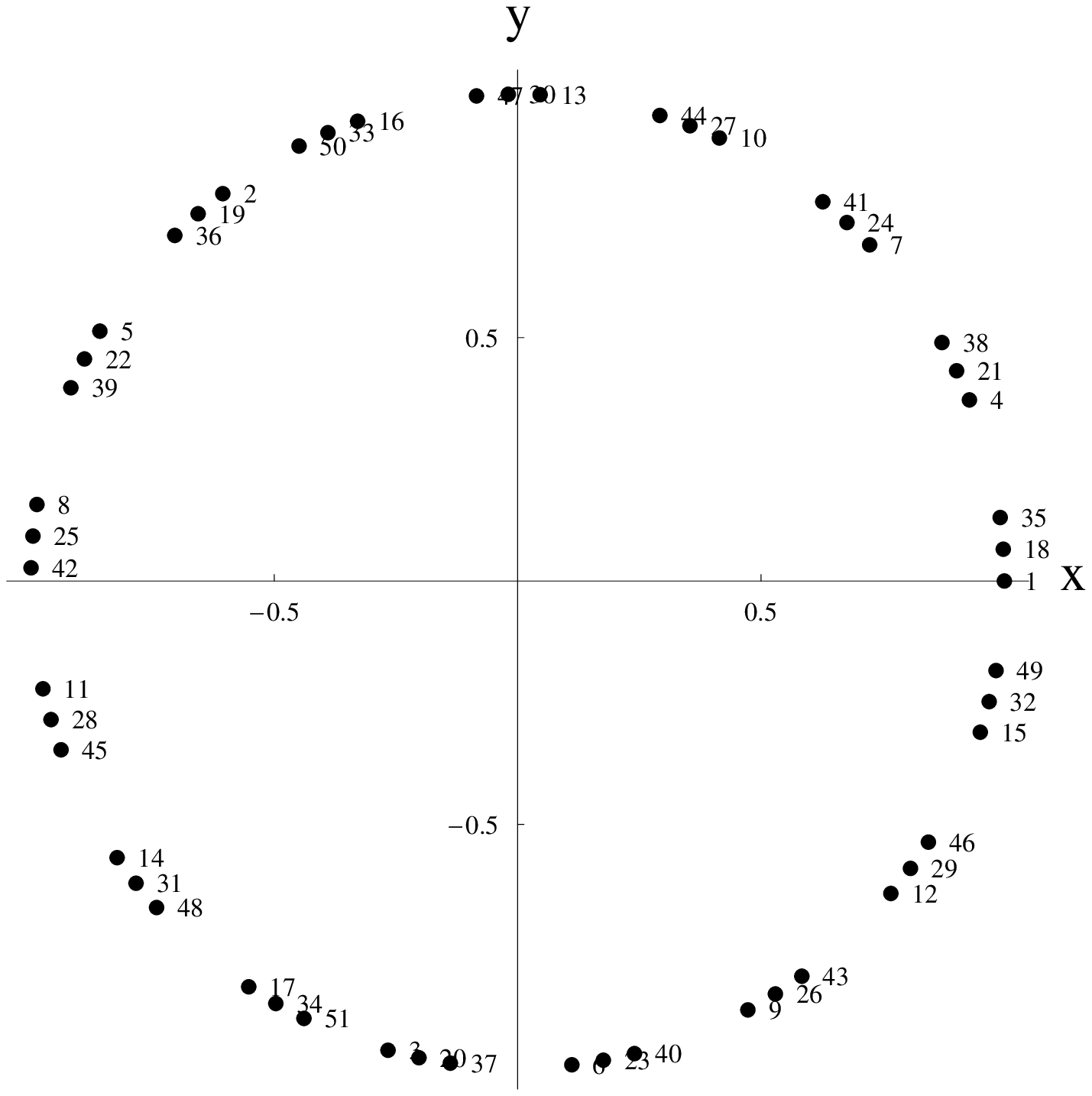}\hskip0.2cm\includegraphics[width=0.29\textwidth]{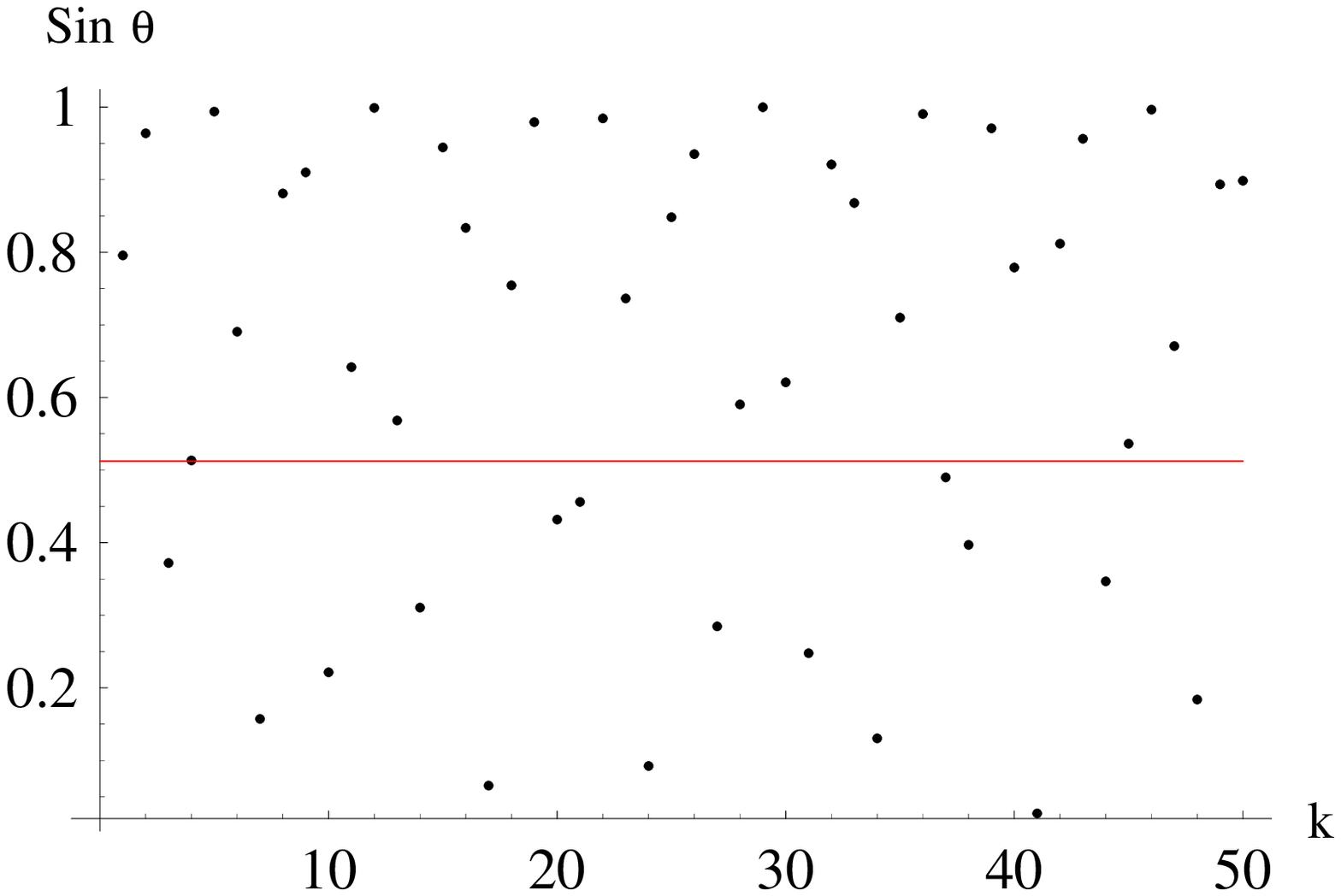}}
\caption{{\bf (a)}: Distribution on the trigonometric circle of the values of $\sin\theta$ and $\cos\theta$ associated with $\theta=k\pi/\sqrt{2}$ for $k=0,..,50$. {\bf (b)}: Distribution of $|\sin\theta|$ against $k$ compared with the value of $\sqrt{\alpha/\beta}=0.512418$, arbitrarily taken from a Gaussian ensemble of $100$ values chosen in the range $[0,1/\sqrt{2}]$.}
\label{esempio}
\end{figure}

Let us momentarily impose the following condition on the interaction time 
\begin{equation}
\label{condn 2}
\cos^2(\lambda t \sqrt{2}) = 1\longrightarrow{\lambda{t}}=k\pi/\sqrt{2}\,\,(k\in\mathbb{Z}). 
\end{equation}
It is clear that this will set 
$E_S\left( \ket{\psi_{eg}}\right) = 
E_S\left( \ket{\psi_{ge}}\right) = 
E_S\left( \ket{\psi_{gg}}\right) = 0$.  
We now make a simple trigonometric observation: As $k\pi/\sqrt{2}$ and $k\pi$ are incommensurable, by letting $k$ grow, the value of $\sin(k \pi / \sqrt{2})$ will fluctuate within the range $[-1,1]$, such that the entire range will eventually be covered.  This is shown pictorially by the trigonometric circle in Fig.~\ref{esempio} (a).


Thus, having set the constraint in Eq.~(\ref{condn 2}), we can still find a value for the rescaled interaction time $\lambda{t}$ such that any additional condition we choose to set on $\sin({\lambda t})$ will also be satisfied.  The only requirement is having a sufficiently large range of integers $k$, {\it i.e.}, a large upper bound on the interaction time $t$~\cite{commentotempo}. 
For instance, if we select $\sin(\lambda t) = 0$, we see that the average ESPP of the cavity modes is zero.
If, on the other hand, $\sin^2(\lambda t) = \alpha/\beta$, we find that condition (\ref{condn for bell state}) is satisfied, so that $E_S\left( \ket{\psi_{ee}}\right) = 1$, and $\left< E_S \right>_f = P\left( \ket{\psi_{ee}}\right) = 2\alpha^2$.
Thus, for general values of $t$, $\left< E_S \right>_f \leq 2\alpha^2$, the upper bound being attained when both conditions~(\ref{condn for bell state}) and~(\ref{condn 2}) are satisfied. In this case, the procedure gives the maximum probability of producing a full-ebit state, and can be said to be optimal.
Note that in this case, a single round of the concentration procedure is all that is necessary to (probabilistically) produce a maximally entangled state from any input pair. 

As an example, in Fig.~\ref{esempio} we show the case of $\alpha=0.253964$, corresponding to $\sqrt{\alpha/\beta}=0.512418$ and $k$ free to span the first $50$ integers. Panel {\bf (a)} clearly shows how these fifty values are uniformly distributed over the trigonometric circle. On the other hand, in panel {\bf (b)} we compare the distribution of $|\sin(k\pi/\sqrt{2})|$ with the chosen value for $\sqrt{\alpha/\beta}$, (which has been extracted from a Gaussian ensemble of values of $\alpha$ lying in the range $[0,1/\sqrt{2}]$). The case is rather favorable as, for $k=4$, $\sin(2\sqrt{2}\pi)=0.513288 \approx \sqrt{\alpha / \beta}$. Clearly, for other values of $\alpha$, larger values of $k$ may be required in order for conditions (\ref{condn for bell state}) and (\ref{condn 2}) to simultaneously hold.


\begin{figure}[t]
\centerline{\includegraphics[width=0.4\textwidth]{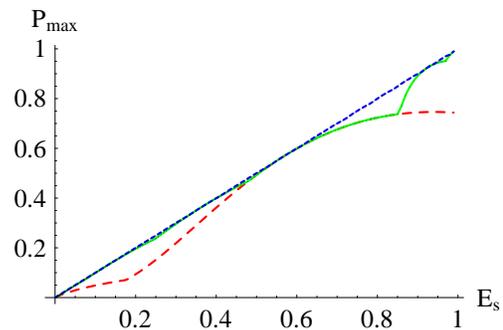}}
\caption{Maximum probability, $P_{max}$, of the entanglement concentration protocol yielding a Bell state,  plotted against the input ESPP ($E_S$) for the case of $\lambda{t}$ bounded by $\pi,\,5\pi$ and $20\pi$ (dashed, solid and dotted line respectively). The larger the upper bound to the rescaled interaction time, the closer the protocol to the ideal performance $P_{max}\equiv{E}_S$, implying conservation of the initial ESPP under application of the concentration protocol~\cite{s. bose et al pra 60 194 (1999)}.} 
\label{fig:fig1}
\end{figure}

However, in practical realizations, the interaction time is limited by the coherence time of the system, which is given by ${\tau}_{cohe}=\min(\tau_{atom},\tau_{field})$, where $\tau_{cavity}$ ($\tau_{atom}$) is the cavity field (excited atomic state) lifetime. Our unitary treatment for entanglement concentration has to be consistent with the requirement $t<\tau_{cohe}$. We therefore change our approach by firstly requiring that condition~(\ref{condn for bell state}) be fulfilled, so that all successful runs produce a Bell state output.  
Secondly, the probability of success is maximized within some upper bound on $\lambda{t}$.  Fig.~\ref{fig:fig1} shows this maximum probability, $P_{max}$, against the ESPP of the input state Eq.~(\ref{non max phiplus}) when the rescaled interaction time $\lambda{t}$ is limited by $\pi,\,5\pi$ and $20\pi$ (dashed, full and dotted line respectively). The implications of these upper bounds in terms of requirements for $\tau_{cohe}$ are discussed in Sec.~\ref{practical}. The analysis of Fig.~\ref{fig:fig1} confirms that a sufficiently large upper bound to the interaction time brings the protocol to optimality. The maximum probability of concentrating entanglement into a full-ebit state becomes equal to $2\alpha^2$, and so the ESPP of the input state is, therefore conserved. Smaller upper bounds will, in general, reduce $P_{max}$. Nevertheless, it is important to stress that even at very short interaction times ({\it i.e} $\lambda{t}<\pi$, dashed line), there are input states $\ket{\Psi_o}_{12}$ allowing for optimal concentration.

The successful implementation of this concentration protocol clearly relies on our previous knowledge of the amplitude $\alpha$ in the initial entangled atomic state $\ket{\Psi_o}_{12}$ in Eq.~(\ref{non max phiplus}). Here, we want to assess the effect of an uncertainty in the value of $\alpha$ on probability of success of the protocol, i.e., the probability of projecting both ancillary atoms onto their excited state. Our assumption is that $\alpha$ follows a Gaussian distribution centered at $\bar{\alpha}$ with standard deviation $\sigma$. This is perfectly reasonable as it is consistent with the protocol we suggested in order to generate input state (\ref{non max phiplus}), where an uncertainty in the velocity of the preparatory atom (usually assumed to follow a Gaussian distribution) may result in some fluctuations in the degree of entanglement between atoms $1$ and $2$. The quantity of interest is the mean success probability $P(\ket{\psi}_{ee})$, as given in Eq.~(\ref{prob}), averaged over the distribution $G(\bar{\alpha},\sigma)=\frac{1}{\sqrt{2 \pi \sigma^2}} {\mbox {exp}}[-\frac{(\alpha-\bar{\alpha})^2}{2\sigma^2}]$.
%
%
%
A numerical analysis reveals that the values of the rescaled interaction time giving the maximum success probability are unaffected by $\sigma$, but the success probability decreases as the uncertainty in $\alpha$ increases, as should be expected.  Clearly, the average entanglement of output states will decrease with increasing $\sigma$.

\subsection{Asymmetric Concentration Scheme}

This protocol is similar to that described above, the difference being that the cavities are prepared in the vacuum state, which eases the implementation of the protocol. After the atoms, prepared in state~(\ref{non max phiplus}), interact with the cavities for a time $t$, the state of the system reads
\begin{eqnarray}
\ket{psi(t)}_{ab12} &=&
[
   -\alpha \sin^2(\lambda t)\ket{11}_{ab} + \beta \ket{00}_{ab}
]
\ket{gg}_{12}
\nonumber
\\
&& -\frac{i}{2}\sin(2\lambda{t})
[
   \ket{01}_{ab}\ket{eg}_{12}
   +
   \ket{10}_{ab}\ket{ge}_{12}]
\nonumber
\\   
&&+
\alpha \cos^2(\lambda t)\ket{00}_{ab}\ket{ee}_{12}
\end{eqnarray}
As opposed to the symmetric scheme, we here require $\alpha>\beta$ and set $\sin^2(\lambda t) ={\beta}/{\alpha}$.  By projecting both atoms onto their ground states, we obtain, with the maximum allowed probability $2\beta^2$, a maximally entangled state of the cavity fields.  Note that we have imposed only one condition on $t$; thus the optimal interaction time will always exist within the limit $\lambda t < \pi/2$.  The required form of the input state can be generated from any input by means of local rotations performed on qubits $1$ and $2$ at properly set Ramsey zones ~\cite{hagley,rauschenbeutel}.

\section{Purification Scheme}

\label{purification}

We now move on to the description of a simple quantum state purification scheme~\cite{bennett1, bennett2} that is able to operate on any Bell-diagonal mixed state. A general remark about the tools used in this Section is due. As we mainly deal with bipartite mixed states of two qubits, we shall quantify quantum correlations in terms of the negativity ${E}_{\cal N}$ \cite{zircone}. This quantity is strictly related to the positive partial transposition (PPT) criterion for the separability of quantum states~\cite{npt}. The partially transposed density matrix $\tilde{\varrho}$ is obtained from any given bipartite quantum state by transposing the variables of only one of the two subsystems. The PPT criterion then 
simply reads $\tilde{\varrho}\ge0$, and is necessary and sufficient for the separability of any bipartite two-qubit quantum state.
The `negativity' ${E}_{\cal N}$~\cite{zircone}, which can be easily determined from the density matrix, is defined as the absolute value of the sum of the negative eigenvalues of $\tilde\varrho$ and directly quantifies the violation of the PPT criterion. It is monotonically related to the logarithmic negativity, which operatively quantifies an upper bound to the {distillable entanglement}.

Let us consider a state of the form

\begin{equation}
\label{general mixed}
{\rho} = 
\sum_{j=\pm}A_{j}\ket{\Phi_{j}}\bra{\Phi_{j}}
+
B_{j}\ket{\Psi_{j}}\bra{\Psi_{j}},
\end{equation}

where $\ket{\Phi_{\pm}}=(1/\sqrt 2)(\ket{00}\pm\ket{11})$ and $\ket{\Psi_{\pm}}=(1/\sqrt 2)(\ket{01}\pm\ket{10})$ are the elements of the Bell basis and $\sum_{j=\pm}(A_j+B_j)=1$. Considering that the partial transposition of a two-qubit density matrix has only one negative eigenvalue~\cite{munro}, it is straightforward to show that if $A_+\geq1/2$, $E_{\cal N}=2A_+-1$. Obviously, ${E}_{\cal N} = 0$ for an equal mixture of Bell states.

\begin{figure}[t]
\centerline{\includegraphics[width=0.4\textwidth]{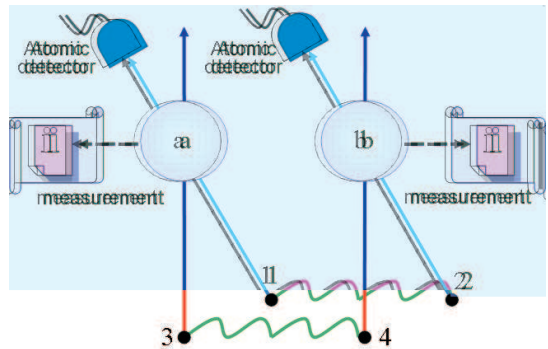}}
\caption{Entanglement purification scheme. Two pairs of flying qubits ($1\&{2}$ and $3\&{4}$) are prepared in the same mixed (Bell-diagonal) state and interact with two cavities ($a$ and $b$ respectively). The atomic-state measurement of pair $1\&{2}$ accompanied by the projection of the cavity fields onto photon-number states realizes a conditional purification of the entangled mixed state shared by qubits $3\&{4}$.}
\label{fig:fig01}
\end{figure}

Let us begin our analysis by assuming the special case of $A_-=B_-=0$ so that, imposing normalization, $A_{+}=1-B_+={\cal P}$. The resulting state reads 
%
\begin{equation}
\label{simple mixed}
{\rho}^{(o)} = 
{\cal P}
\ket{\Phi_{+}}\bra{\Phi_{+}}
+
\left(1-{\cal P}\right)
\ket{\Psi_{+}}\bra{\Psi_{+}},
\end{equation}
for which ${E}_{\cal N}^{(0)} =2{\cal P} -1$. The task of our study is the purification of $\rho_o$ by using two replicas of the state and an ancillary system. In our scheme, the replicas of $\rho_o$ are encoded in the state of qubits $1\,\&\,2$ and $3\,\&\,4$, while the ancillary system is embodied by two cavities (labeled $a$ and $b$) analogous to those considered in the previous Section. In order to illustrate the main features of our protocol, we momentarily assume that the cavities are prepared in an $n$-photon state, with $n \gg 1$. We will relax this assumption later on. The initial state of the system is thus $\rho_{o,12}\otimes\rho_{o,34}\otimes\ket{nn}_{ab}\bra{nn}$.

As shown in Fig.~\ref{fig:fig01}, one atom from each pair passes through each cavity (qubits $1\,\&\,3$ pass through $a$ while qubits $2\,\&\,4$ pass through $b$) and the interaction time $t$ is selected so that $\sqrt{n}\lambda{t}={\pi}/{2}$. In order to understand the main features of the scheme we propose, it is enough to consider just the dynamics within cavity $a$, where, by neglecting the small difference between the effective Rabi frequencies $\sqrt{n}\lambda{t}$ and $\sqrt{n+1}\lambda{t}$, the following transformations occur:

\begin{equation}
\begin{aligned}
\ket{ee}_{13}\ket{n}_a & \rightarrow  \ket{gg}_{13}\ket{n+2}_a,\hskip0.2cm\ket{eg}_{13}\ket{n}_a \rightarrow  \ket{ge}_{13}\ket{n}_a, \\
\ket{gg}_{13}\ket{n}_a & \rightarrow  \ket{ee}_{13}\ket{n-2}_a,\hskip0.2cm\ket{ge}_{13}\ket{n}_a  \rightarrow  \ket{eg}_{13}\ket{n}_a.
\end{aligned}
\end{equation}

The cavity fields are now projected onto $\ket{nn}_{ab}$ with a success probability $[{\cal P}^2 + (1-{\cal P})^2]/2$. Only those terms in which both atoms passing through each cavity are in different states survive the projective measurements.  Thus, if both atom pairs are initially in the same Bell state, the output will be a GHZ-like state of four atoms, while cross terms will be annihilated;






\begin{equation}
\label{uso}
\begin{aligned}
\ket{\Phi_+}_{12}\ket{\Phi_+}_{34}
 &\rightarrow 
\ket{\Phi^2}_{1234} = 
\frac{1}{\sqrt{2}}
(\ket{ggee} + \ket{eegg})_{1234},\\
\ket{\Phi_+}_{12}\ket{\Psi_+}_{34}&\rightarrow  0,\,\,\ket{\Psi_+}_{12}\ket{\Phi_+}_{34}\rightarrow  0,\\
\ket{\Psi_+}_{12}\ket{\Psi_+}_{34}&\rightarrow  
\ket{\Psi^2}_{1234} =
\frac{1}{\sqrt{2}}
(\ket{geeg} + \ket{egge})_{1234}.
\end{aligned}
\end{equation}

The effect of the ancillary systems embodied by the cavity fields is to {\it filter} the right form of correlations among the qubit pairs, leaving us with the mixed state
\begin{equation}
\label{4 party mixed out}
{\rho}^{(1)}_{1234} = 
\frac
{
   {\cal P}^2 \ket{\Phi^2}_{1234}\bra{\Phi^2}
   +
   \left(1-{\cal P}\right) ^2
   \ket{\Psi^2}_{1234}\bra{\Psi^2}
}
{{\cal P}^2 + \left( 1-{\cal P} \right) ^2}.
\end{equation}
  
If we now rotate atom 1 about its {\em z}-axis by angle $\pi$, then use a third cavity to perform the same filtering operation on atoms 1 and 2, we will obtain, with probability 
$
\frac{{\cal P}^4}
{2[{\cal P}^2+(1-{\cal P})^2]^2}
$
 the pure GHZ-like state
\begin{equation}
\ket{\psi}_{1234}
=
\frac{1}{\sqrt{2}}
\left(
   \ket{egee} + \ket{gegg}
\right)_{1234}
\end{equation}
This would, however, require a significantly more complicated experimental setup.  Instead of this final filtering stage, we can use an iterative procedure to produce maximally entangled states of arbitrary purity. 
 
We convert (\ref{4 party mixed out}) to a mixture of bipartite entangled states by measuring atoms 1 and 2 in the $\{\ket{\pm}_{j}\}$ ($j=1,2$) basis, where $\ket{\pm}_j=({1}/{\sqrt{2}})(\ket{e}_j \pm \ket{g}_j)$, which conditionally produces the outputs 






\begin{equation}
\label{ideal simple mixed out}
{\rho}^{(1)}_{\pm,34} = 
\frac
{
   {\cal P}^2 \ket{\Phi_{\pm}}_{34}\bra{\Phi_{\pm}}
   +
   \left(1-{\cal P}\right) ^2
   \ket{\Psi_{\pm}}_{34}\bra{\Psi_{\pm}}
}
{{\cal P}^2 + \left( 1-{\cal P} \right) ^2}.
\end{equation}
The state ${\rho}^{(1)}_{+,34}$ is obtained when atoms 1 and 2 are projected onto the same state, and ${\rho}^{(1)}_{-,34}$ otherwise. It is possible to interconvert ${\rho}_+$ and ${\rho}_-$ by performing a $z$-rotation of angle $\pi$ on one of the atoms. This is unnecessary, however, as $\ket{\Phi_{-}}$ behaves similarly to $\ket{\Phi_{+}}$ under the transformations described above: When applied, the protocol will annihilate $\ket{\Phi_-}\ket{\Psi_{+}}$ terms and preserve $\ket{\Phi_-}\ket{\Phi_{+}}$. Likewise, $\ket{\Psi_{-}}$ behaves analogously to $\ket{\Psi_{+}}$.

Therefore, if ${\cal P} > \hf$, we can see that we have increased the relative proportion of the state $\ket{\Phi_{\pm}}$, and hence its purity. Correspondingly, $E_{\cal N}$ has increased  from $E_{{\cal N}}^{(0)}=2{\cal P}-1$ to $E_{{\cal N}}^{(1)}=\frac{2{\cal P}-1}{{\cal P}^2 + (1-{\cal P})^2}$. By iterating this procedure, it is possible to produce a state of arbitrary purity. By using $q$ replicas of $\rho_o$, the purification protocol produces the final state
\begin{equation}
\label{ideal simple mixed out q}
{\rho}^{(q)}_{34} = 
\frac
{
   {\cal P}^q \ket{\Phi_{+}}_{34}\bra{\Phi_{+}}
   +
   \left(1-{\cal P}\right) ^q
   \ket{\Psi_{+}}_{34}\bra{\Psi_{+}}
}
{{\cal P}^q + \left( 1-{\cal P} \right) ^q}
\end{equation}
with a success probability $({\cal P}^q + (1-{\cal P})^q)/2^q$.

\begin{figure}[b]
\psfig{figure=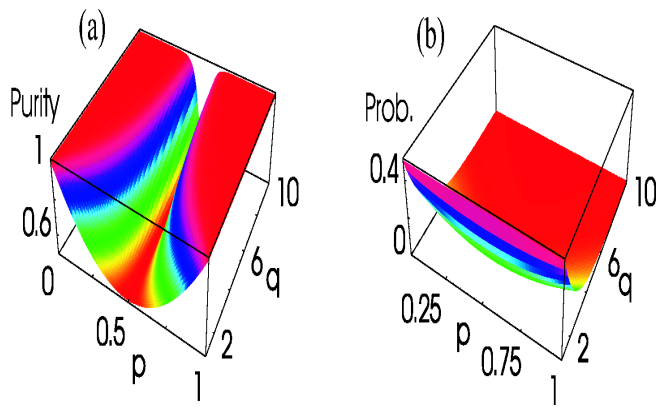,width=9cm,height=5.5cm}
\caption{{\bf (a)}: Purity of the evolved state in the purification protocol against the initial purity ${\cal P}$ and iteration number $q$. {\bf (b)}: Probability of purification against ${\cal P}$ and $q$.}
\label{purifica}
\end{figure}

The evolution of the input state towards a maximally entangled pure state is tracked, in terms of purity and probability of purification, in Fig.~\ref{purifica}. Obviously, the probability of success tends rapidly to zero as the number of iterations performed increases to infinity. Although our success probability is less than that of Bennett {\em et al.,}~\cite{bennett1}, it matches that of Pan {\em et al.}~\cite{J-W. Pan Nature 423 417 (2003)}.

Mixed states of the form 
$
{\cal P'}
\ket{\Phi_{+}}\bra{\Phi_{+}}
+
\left(1-{\cal P'}\right)
\ket{\Phi_{-}}\bra{\Phi_{-}}
$
and
$
{\cal P'}
\ket{\Psi_{+}}\bra{\Psi_{+}}
+
\left(1-{\cal P'}\right)
\ket{\Psi_{-}}\bra{\Psi_{-}}
$
can be purified by means of local conversion into the form of Eq.~(\ref{simple mixed}) (operated through single-qubit rotations) and the application of the scheme depicted above.

In the following two Subsections, we provide two examples of applications of the purification protocol we have described so far. The objects of our interest are Werner states and MEMS.

\subsection{Purification of a Werner state}

Let us consider the Werner state
\begin{equation}
\label{werner state}
{\rho}_{W}^{} = 
{\cal P}\ket{\Phi_{+}}\bra{\Phi_{+}}
+
{\cal Q}
\left(\openone-
   \ket{\Phi_{+}}\bra{\Phi_{+}}
\right),
\end{equation}
where ${\cal Q} = (1-{\cal P})/3$.  Such states are (for general ${\cal P}$) mixed, and entangled for ${\cal P} > \frac{1}{2}$, as witnessed by the negative partial transposition criterion.

In approaching the purification of Eq.~(\ref{werner state}), some complications arise due to the fact that cross terms between atom pairs in the states $\ket{\Phi_{+}}$ and $\ket{\Phi_{-}}$ would not be annihilated. Indeed, after the first round of purification, the output state (conditioned on qubits $1$ and $2$ being projected onto the same state) reads


\begin{equation}
\label{werner pure rd 1}
\begin{aligned}
{\rho}_{W}^{(1)}=&{\cal B} 
   \left[
      \left( {\cal P}^2 + {\cal Q}^2 \right) \ket{\Phi_{+}}\bra{\Phi_{+}}   + 
      2{\cal P}{\cal Q} 
      \ket{\Phi_{-}}\bra{\Phi_{-}}
   \right.
   \\
&   +
   \left.
   4{\cal Q}^2
   \left(
      \ket{\Psi_{+}}\bra{\Psi_{+}}
      +
      \ket{\Psi_{-}}\bra{\Psi_{-}}
   \right)
   \right]
\end{aligned}
\end{equation}


with ${\cal B}={{\cal P}^2+2{\cal P}{\cal Q}+5{\cal Q}^2}$. 


This state does not exhibit any great improvement, in terms of purity, with respect to $\rho_{W}$: Indeed, repeating the procedure using states (\ref{werner state}) and (\ref{werner pure rd 1}) as inputs actually causes a decrease in negativity, below that of $\rho_W$.

In order to conduct further purification, we prepare two atom pairs in state (\ref{werner pure rd 1}), and subject each atom to a $\frac{\pi}{2}$ rotation around its $x$-axis.  This will effect the transformations 
$\ket{\Phi_+}$ to $\ket{\Phi_+}$, 
$\ket{\Phi_-}$ to $\ket{\Psi_+}$,  
$\ket{\Psi_+}$ to $\ket{\Phi_-}$ and
$\ket{\Psi_-}$ to $\ket{\Psi_-}$.  
Using these two modified pairs as the inputs for our protocol, we now obtain the output
\begin{widetext}
\begin{equation}
\label{werner pure rd 2}
{\rho}_{W}^{(2)} = 
\frac
{
   \left[
      {\cal P}^4 
      + 2{\cal P}^2{\cal Q}^2
      + 5{\cal Q}^4
   \right]
   \ket{\Phi_{+}}\bra{\Phi_{+}}
   + 
   4\left[
      {\cal P}^2
      {\cal Q}^2
      +
      {\cal Q}^4
   \right]
   (\ket{\Phi_{-}}\bra{\Phi_{-}}+\ket{\Psi_{+}}\bra{\Psi_{+}})
   +
   8{\cal P}
   {\cal Q}^3
   \ket{\Psi_{-}}\bra{\Psi_{-}}
}
{
   {\cal P}^4
   +
   10{\cal P}^2
   {\cal Q}^2
   +
   8{\cal P}
   {\cal Q}^3
   +
   13
   {\cal Q}^4
}
\end{equation}
\end{widetext}
which is purified with respect to $\rho_{W}$ (we can see that it is pure to first order in ${\cal Q}$).  As before, an arbitrarily pure state can be obtained by iteration, but the cumulative probability of success decreases rapidly with the number of steps.

\subsection{Purification of MEMS}

We now approach MEMS as the second non-trivial example of the applicability of our purification protocol. MEMS are states possessing the largest achievable mixedness for a given degree of entanglement $ E_{\cal N}\ge{0}$~\cite{munro}.
A parameterization of MEMS is critically dependent on the
chosen measures of entanglement and purity. If $E_{\cal N}$ is taken to quantify entanglement and the linearized entropy $S_{l}=(4/3)(1-\mbox{Tr}\rho^2_{})$ is used as the measure of the purity of a state, MEMS are a single-parameter family of states given by
\begin{equation}
\label{memsstate}
\rho_{mems}=
\begin{pmatrix}
\frac{1+\sqrt{1+3g^2}}{6}&0&0&\frac{g}{2}\\
0&\frac{2-\sqrt{1+3g^2}}{3}&0&0\\
0&0&0&0\\
\frac{g}{2}&0&0&\frac{1+\sqrt{1+3g^2}}{6}
\end{pmatrix}
\end{equation}
with $0\le{g}\le{1}$~\cite{munro}. MEMS have been theoretically characterized and schemes for their generation in cavity-QED and circuit-QED have been proposed~\cite{mems}. Based on a ``Procrustean" method, they have also been experimentally produced in an all-optical setup~\cite{kwiatmems}.  In order to apply the purification protocol, we rewrite $\rho_{mems,12}\otimes{\rho_{mems,34}}$ in the Bell basis, let it interact with the cavity fields and then exploit the transformations in Eqs.~(\ref{uso}) and the analogous transformations valid for the remaining tensorial product of two input Bell states. After the projection of qubits $1$ and $2$ onto the $\hat{\sigma}_x$ eigenbasis, we are left with the reduced state of qubits $3$ and $4$

\begin{equation}
\label{memspurificata}
\rho^{(1)}_{mems,34}=\left[
\begin{matrix}
\frac{1}{2}&0&0&\frac{5}{6}-\frac{2}{3\sqrt{1+3g^2}}\\
0&0&0&0\\
0&0&0&0\\
\frac{5}{6}-\frac{2}{3\sqrt{1+3g^2}}&0&0&\frac{1}{2}
\end{matrix}
\right].
\end{equation}

\begin{figure}[t]
\centerline{\includegraphics[width=0.4\textwidth]{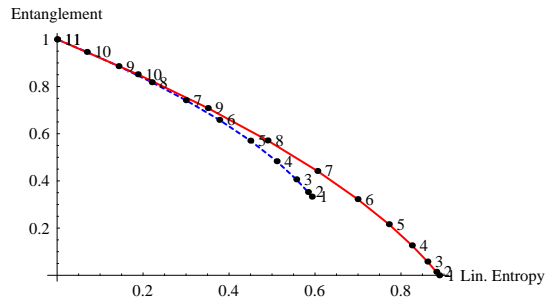}}
\caption{Comparison, in terms of linearized entropy $S_l$ and negativity $E_{\cal N}$, between $\rho_{mems,34}$ and $\rho^{(1)}_{mems,34}$, obtained after the purification protocol. The solid (dashed) line shows the input (purified) state. The numbered points on each curve identify the position of the states associated with eleven values of $g=0,0.1,..,1$. }
\label{purificomems}
\end{figure}

This state is purified with respect to the original MEMS, and has, for a set value of $g$, a larger negativity. The quantitative comparison between the purified and input state is shown in Fig.~\ref{purificomems}, where the states corresponding to eleven discrete values of $g\in[0,1]$ (equally spaced with steps of $0.1$) are plotted in the negativity-linear entropy plane. We recall that, by definition, $S_l=0$ for pure states and $S_l=1$ for a totally mixed state. Obviously, in agreement with the interpretation of MEMS as boundary states to physically achievable entangled mixed states, the purified states describe a trajectory in the $E_{\cal N}-S_l$ plane below the extremal curve traced by $\rho_{mems,34}$.

Evidently, for a set value of $g$, the corresponding $\rho^{(1)}_{mems,34}$ has a smaller $S_l$, meaning a larger purity, and a larger value of negativity, than $\rho_{mems}$.

\subsection{Simplified purification protocol}

Producing high photon-number Fock states, upon which the protocol discussed so far is based, is rather difficult. Here, in order to reduce the complexity of implementing the scheme, we briefly discuss a simplified version
in which the two cavities are initially prepared in single-photon states. The first pair of atoms interacts with the cavities for a time $t$ chosen so that $\lambda t = (m_1 + \hf)\pi $, with $m_1$ an integer. Analogously, the interaction time for the second pair, $t'$, is such that $\sqrt{2}\lambda t'  = (m_2 + \hf)\pi $.  The evolved states involve cavity field modes with up to three photons. 

























%

By projecting the cavity modes onto the single-photon state, we realize effective transformations that read, in the computational atomic basis
\begin{equation}
\begin{aligned}
\ket{eg}_{13} & \rightarrow 
   \cos{\vartheta_1}\cos{\vartheta_2} \ket{eg}_{13} 
   -i(-1)^{m_2}
   \sin{\vartheta_1}\ket{ge}_{13},
\\
\ket{ge}_{13} & \rightarrow  
   \sin{\vartheta_2}\ket{eg}_{13},
\end{aligned}
\end{equation}
where $\vartheta_1 =\pi\sqrt{2}\left(m_1 + \hf\right)$ and $\vartheta_2 =\frac{\pi}{ \sqrt{2}}\left(m_2 + \hf\right)$.
All terms in the input states involving $\ket{ee}_{13}$ and $\ket{gg}_{13}$ are annihilated by the field-state measurements that, again, act as filters. By executing this procedure on two atomic pairs, each prepared in state~(\ref{simple mixed}), we obtain (if atoms 1 and 2 are both projected onto $\ket{+}$) the output
\begin{equation}
\label{nonideal simple mixed out}
\hat{\rho}_{s}^{(1)}
=
\frac
{
   {\cal P}^2 N_{\Phi}\ket{\Phi'}\bra{\Phi'}
   +
   \left(1-{\cal P}\right) ^2 N_{\Psi}
   \ket{\Psi'}\bra{\Psi'}
}
{{\cal P}^2 N_{\Phi} + \left( 1-{\cal P} \right)^2 N_{\Psi}}
\end{equation}
where
\begin{equation}
\label{not quite bell states}
\begin{aligned}
\ket{\Phi'} &=
   \frac{1}{\sqrt{N_{\Phi}}}
   [
      \sin^2{\vartheta_2} \ket{gg}
      -
      \sin^2{\vartheta_1} \ket{ee}
      +
      \left(
         \cos{\vartheta_1} \cos{\vartheta_2}
      \right)^2
      \ket{gg}\\
&      -i(-1)^{m_2}
      \cos{\vartheta_1} \cos{\vartheta_2} \sin{\vartheta_1}
      \left(
         \ket{eg} + \ket{ge}
      \right)
   ],
\\
\ket{\Psi'} &=
   \frac{1}{\sqrt{N_{\Psi}}}
   [
      -i(-1)^{m_2}\sin{\vartheta_1} \sin{\vartheta_2}
      \left(
         \ket{eg} + \ket{ge}
      \right)\\
& +
       \cos{\vartheta_1} \sin{(2\vartheta_2)}
      \ket{gg}
   ]
\end{aligned}
\end{equation}
and $N_{\Psi},\,N_{\Phi}$ are normalization factors. By choosing $m_1$ and $m_2$ such that $\cos{\vartheta_1} \approx \cos{\vartheta_2} \approx 0$, the states (\ref{not quite bell states}) reduce to Bell states, and we recover an approximation of the ideal $n$-photon output, (\ref{ideal simple mixed out}).
This can be achieved with rescaled interaction times of only a few Rabi cycles, which is advantageous from an experimental point of view. For instance, setting $m_2 =m_1+1=3$, we obtain $\left| \braket{\Phi_-}{\Phi'}\right|^2 = 1-9.8\times10^{-6}$.



%



















So far, we have overlooked the difficulty of projecting cavity modes onto the single-photon state.  In our particular situation, this is reduced to the simpler problem of distinguishing a single-photon state from zero-, two- and three-photon states.  In fact, the latter of these can be effectively removed from the mixture by selecting $m_2$ such that $\sin{\sqrt{2}\vartheta_2} \approx 0$ (compatible with $\sin\vartheta_2\simeq{1}$ for $m_2=7$). In order to accomplish the measurement, we pass an atom in its ground state through the cavity.  The interaction time, $t''$, is selected so that $\sin{\left( \lambda t'' \sqrt{2}\right)} = 0$ and $\cos{(\lambda t'')} \approx 0$.  If the cavity was initially in a superposition of zero-, one- and two-photon states, the interaction will produce the transformation
\begin{equation}
\begin{aligned}
(c_0\ket{0} + c_1\ket{1} + c_2\ket{2})\ket{g}{\rightarrow}{}
   c_0\ket{0}\ket{g}
   -i
   c_1\ket{0}\ket{e}   +
   c_2\ket{2}\ket{g}
\end{aligned}
\end{equation}
If, on observing the atom, we find it to be in its excited state, we can conclude that the cavity must have been in a single-photon state.

\section{Practical considerations}

\label{practical}

In this Section, we discuss potential experimental settings for the implementation of our entanglement concentration and purification schemes. As already mentioned, we will explicitly address microwave cavity- and circuit-QED set-ups, which are two of the most promising scenarios for quantum information processing and communication, as stated in the Los Alamos roadmap~\cite{losalamos}.

Microwave cavity-QED uses a mature technology based on the interaction between neutral Rydberg atoms of large dipole moments and high-finesse microwave cavities~\cite{hagley,rauschenbeutel,haroche}. Paradigmatically, we follow the proposal of Kuhr {\em et al.}~\cite{kuhr}, of superconducting cavities sustaining a field mode of frequency $\omega_f/2\pi=51$ GHz, length $L\simeq27$ mm and energy damping time $\tau_{field}\simeq130$ ms. This corresponds to a cavity finesse $F=\pi{c}\,\tau_{field}/{L}\sim5\times10^{9}$, where $c$ is the speed of light. As previously mentioned, the coherence time of the overall system depends also on the damping time of the excited state of the flying qubits (each encoded in circular states with principle quantum numbers $50$ and $51$ of a Rydberg atom), which is comparable to $\tau_{field}$. The atomic transition frequency ($\omega_0=51.09$ GHz) can be tuned in and out of resonance with the cavity field by means of a d.c. Stark shift induced by a static electric field parallel to the cavity axis (this field also helps preserve the circularity of the atomic states). This technique can be used to control the interaction time between each flying atom and the corresponding cavity field. The interaction time is also dependent on the atomic velocity across the resonator, which is usually in the range of hundreds of m.s$^{-1}$, with an experimental accuracy well-within $5-10\%$. The effect of this uncertain atomic velocity on the concentration protocol is discussed in Section~\ref{concentration}. The large atomic dipole moment ($\sim10\,{e}a_0$ where $e$ is the electron charge and $a_0$ is the Bohr radius) allows for the achievement of a strong coupling regime characterized by $\lambda=2\pi\times50$ kHz. This value guarantees the possibility of performing $100$ Rabi floppings ({\it i.e.} $\lambda{t}\simeq{100}\pi$) within $\tau_{cohe}$~\cite{kuhr}, putting the requirement for a long upper bound on the rescaled interaction time (as stated for the concentration protocol) within the present state of the art. Atomic state detection, necessary for both the concentration and the purification scheme, can be accurately performed with state-selective field-ionization channeltrons (each having a detection-error between $10$ and $13\%$). With the new generation of high-finesse cavities addressed above, the atomic injection is performed through large apertures, allowing for the screening of the flying qubits from coherence-destroying stray fields at the surface of the cavity mirrors. 

In a circuit-QED setup, flying atoms are replaced by static superconducting qubits embodied by SQUIDs working in the charge regime at the degeneracy point (to remove, to first order in the single-Cooper pair charge $2e$, the detrimental effect of low-frequency noise induced by background impurities)~\cite{schon,schoelkopf}. The qubit is integrated, via conventional optical lithography, in a full-wave, on-chip, coplanar waveguide cavity with resonant frequency $\omega_{field}=5.7$ GHz (the microstrip resonator), located at a voltage antinode of the sustained field mode and capacitively coupled to it. The stripline is a quasi-unidimensional structure with a very small transversal dimension that reduces the effective volume of the cavity field and enhances the coupling rate with the qubit. This, together with the effective dipole moment of the SQUID qubit ($\sim{2}\times10^{4}ea_0$) gives rise to the ratio $\lambda/\omega_{field}\simeq{2\%}$. The energy damping time of the stripline is conservatively assumed to be $\sim1\mu$s, giving $\lambda\tau_{field}=100$, which would, in principle, allow for a large number of coherent Rabi floppings within the cavity lifetime (experimental evidence puts the qubit damping rate in the range of $2\mu$s), in complete analogy with the cavity-QED setting described above. A detailed derivation of the qubit-stripline coupling Hamiltonian and the resulting coupling strength can be found in Paternostro {\it et al.}~\cite{schoelkopf}.

The transition energy of the superconducting qubit can be adjusted through an external magnetic flux that modulates the Josephson energy of the SQUID~\cite{schon} in such a way that the qubit can be easily put in the strong resonant or dispersive regime with the field. This tuning ability is the basis of the experimentally demonstrated non-demolition measurement of the qubit state, through spectroscopic resolution of the field's frequency-pulling effect~\cite{schoelkopf}. Therefore, the projection of a qubit required by our schemes can be reliably implemented by using an ancillary cavity field mode, differing in frequency or polarization with respect to the mode singled out for the realization of our protocols, along the lines depicted in~\cite{schoelkopf}. On the other hand, the same technique can be used in order to perform a photon-number-resolving cavity field measurement: The qubit spectrum exhibits a peak that is shifted, with respect to the empty cavity situation, by a quantity depending on the population of the field mode.

The integration of more than a single SQUID qubit in the stripline, in a way that avoids (inductive) cross-talk between the qubits, is achievable, the main difficulty being the necessity of separating each qubit's gate voltage~\cite{schon} to properly set each working point. As opposed to cavity-QED, however, the entanglement between superconducting qubits in circuit-QED has yet to be experimentally demonstrated, despite promising steps having been performed along these lines~\cite{schoelkopfnotes}. 

For the sake of completeness, we here mention that a scheme for the preparation of MEMS has been suggested, both in cavity- and circuit-QED, in Ref.~\cite{mems}. We refer you to these works for an extensive account of the details necessary for this procedure.

\section{Conclusion}

\label{conclusions}

We have presented two schemes for entanglement concentration and purification for qubit systems. Our protocols are explicitly designed in settings involving qubits interacting with ancillary cavity fields, which can be implemented in cavity- and circuit-QED set-ups. Our entanglement concentration scheme conserves ESPP, therefore being optimal, and achieves the 
maximum theoretical probability of producing a maximally entangled state of two qubits. The state purification protocol 
can be iterated so as to return a state of arbitrary purity, although the success probability decreases exponentially as perfect purity is approached. 


The required resources are modest. We need cavities prepared in zero- and single-photon states, which have been experimentally demonstrated, and the ability to perform single-atom rotations and measurements, which is possible with high accuracy and large detection efficiencies. In addition, we require cavity-field measurements, which can be achieved in an indirect way: We have addressed strategies for this step in both of the proposed experimental scenarios. 



The mechanisms that constitute the main sources of imperfection and error have been quantitatively assessed to provide a comprehensive analysis of the protocols at hand.  We hope that the evidence for state-of-the-art implementability of our schemes and the astonishing improvements in the control of microwave-based quantum technology will pave the way to the realization of entanglement concentration and purification in cavity- and circuit-QED systems.

\acknowledgments

We acknowledge support from the UK EPSRC, The Leverhulme Trust (ECF/40157) and DEL.


\begin{thebibliography}{9}
\bibitem{commentotempo} In principle, the condition on the rescaled interaction time sets a constraint only on the product $\lambda{t}$. Therefore, Eq.~(\ref{condn for bell state}) can be satisfied for a proper choice of either $\lambda$ or $t$. However, the interaction strength $\lambda$ is determined by the specific setup used in order to implement the scheme, leaving us with the freedom of chosing a value of $t$ (within the coherence time of a given setup) that suits Eq.~(\ref{condn for bell state}) with the required accuracy.
\end{thebibliography}
\end{document}